\documentclass{emulateapj}
\usepackage{xspace}
\usepackage{amsmath}
\usepackage{hyperref}
\usepackage{framed} 
\usepackage{txfonts}
\usepackage{epstopdf} 
\usepackage{color}

\usepackage{ulem}

\def\Ht{{\rm H}}
\def\H2{{{\rm H}_2}}
\def\HI{{\rm H\,I}}

\def\Msun{\, M_{\odot}}

\def\Sgas{\Sigma_{\rm H}}
\def\Sntr{\Sigma_{\HI+\H2}}

\def\Smol{\Sigma_\H2}
\def\Shi{\Sigma_\HI}

\def\D{D_{\rm MW}}
\def\U{U_{\rm MW}}

\def\dim#1{\mbox{\,#1}}

\begin{document}

\title{Line Overlap and Self-Shielding of Molecular Hydrogen in Galaxies}

\author{Nickolay Y.\
  Gnedin\altaffilmark{1,2,3} and Bruce T.\ Draine\altaffilmark{4}}  
\altaffiltext{1}{Particle Astrophysics Center, 
Fermi National Accelerator Laboratory, Batavia, IL 60510, USA; gnedin@fnal.gov}
\altaffiltext{2}{Kavli Institute for Cosmological Physics, The University of Chicago, Chicago, IL 60637 USA;
  andrey@oddjob.uchicago.edu} 
\altaffiltext{3}{Department of Astronomy \& Astrophysics, The
  University of Chicago, Chicago, IL 60637 USA} 
\altaffiltext{4}{Princeton University Observatory, Princeton, NJ 08544-1001; draine@astro.princeton.edu}

\begin{abstract}
The effect of line overlap in the Lyman and Werner bands, often ignored in galactic studies of the atomic-to-molecular transition, greatly enhances molecular hydrogen self-shielding in low metallicity environments, and dominates over dust shielding for metallicities below about 10\% solar. We implement that effect in cosmological hydrodynamics simulations with an empirical model, calibrated against the observational data, and provide fitting formulae for the molecular hydrogen fraction as a function of gas density on various spatial scales and in environments with varied dust abundance and interstellar radiation field. We find that line overlap, while important for detailed radiative transfer in the Lyman and Werner bands, has only
a minor effect on star formation on galactic scales, which, to a much larger degree, is regulated by stellar feedback.
\end{abstract}


\section{Introduction}
\label{sec:intro}

In the last several years an important advance has been made in understanding star formation on galactic scales. Both local \citep{ism:lwbb08,sfr:blwb08,sfr:bljo11,sfr:blwb11,sfr:lbbb12,sfr:lwss13} and intermediate redshift \citep{sfr:gtgs10,sfr:dbwd10,sfr:tngc13} observational studies find that the star formation rate surface density on kpc scales correlates well, and approximately linearly, with the surface density of molecular gas.

Hence, from a theoretical perspective, tracing the formation of molecular hydrogen in cosmological and galactic-scale simulations is a pre-requisite for modeling star formation and the subsequent stellar feedback. Several groups have recently implemented models of atomic-to-molecular gas transition in cosmological simulations codes and explored their predictions \citep{ism:ppw06,ism:rk08,ng:gtk09,ism:pp09,ng:gk11,sfr:cqgs12,sfr:kkms12,sfr:kmk13,sfr:tnj14}. However, all these models either completely ignored the self-shielding of molecular hydrogen, or included it in the approximation where each of the Lyman and Werner band absorption lines is treated as isolated - in that limit shielding by cosmic dust is important, and the characteristic column density of the atomic-to-molecular transition scales inversely proportional to the dust abundance. 

Such an approximation is appropriate for sufficiently low column densities of molecular gas. However, for $N_\H2 \gg 10^{21}\dim{cm}^{-2}$ the damping wings of individual Lyman and Werner band absorption lines begin to overlap \citep{sw67,bd77,ism:db96}. The process of line overlap, when applied to a molecular interstellar medium (ISM) with supersonic turbulence, results in a significant enhancement in the role of self-shielding; in sufficiently low metallicity environments self-shielding may, in fact, dominate over the dust shielding, as we demonstrate below.

In complex environments of realistic galaxies there are many other physical processes that affect the chemical and dynamical state of the ISM; molecular hydrogen self-shielding may or may not be an important process in such environments, but a study is warranted to explore its role. This paper aims at addressing a part of this question - namely, the role of line overlap in the transition from atomic to molecular hydrogen in galaxies.

\section{Self-shielding of Molecular Hydrogen}
\label{sec:self}

The molecular hydrogen photodissociation rate $\zeta_{\rm pd}$ can be
written
\begin{equation}
\zeta_{\rm pd} = S_{\H2}\times S_{\rm dust}\times \zeta_{\rm pd}^0
\end{equation}
where $\zeta_{\rm pd}^0$ is the ``free space'' rate, $S_{\rm dust}$ is the
reduction in the rate due to shielding by dust, and $S_{\H2}$ is the
reduction factor due to $\H2$ self-shielding.
Self-shielding of molecular hydrogen has been studied extensively  since the pioneering work of \citet{sw67}. A commonly used formula that conveniently parametrizes the self-shielding effects for $\H2$ with (local) one-dimensional
velocity dispersion $\sigma_v$ was given by \citet{ism:db96},
\begin{equation}
S_{\H2} = \frac{0.965}{(1+x/b_5)^2} + \frac{0.035}{\sqrt{1+x}}\exp\left(-\frac{\sqrt{1+x}}{1180}\right),
\label{eq:shfac}
\end{equation}
where $x \equiv N_\H2/5\times10^{14}\dim{cm}^{-2}$, and $b_5 \equiv b/\dim{km/s}$, with $b\equiv\sqrt{2}\sigma_v$.
Here $N_\H2$ is the column density of $\H2$ between the point of interest
and the sources of $111-91.2$nm radiation that can dissociate $\H2$ with
$v=0$ and $J=0,1,2$.
The accuracy of this shielding function has been confirmed
by recent work \citep[][see their Fig.\ 5]{Sternberg+LePetit+Roueff+LeBourlot_2014}, who also explicitly accounted for line overlap in the Lyman and Werner bands.

Equation (\ref{eq:shfac}) is suitable for the idealized case of a uniform slab of gas with no internal motions. Real molecular clouds are, however, supersonically turbulent on scales above the sonic length, $l_s\la 1\dim{pc}$. If we consider two fluid elements in the molecular cloud separated by a large distance $L\gg l_s$, the velocity difference $\Delta v \sim b (L/l_s)^\gamma$ \citep[with $\gamma\approx0.5$ from the Larson Law;][]{sfr:mo07} between them would be large, much larger than the local Doppler width $b$ of each Lyman and Werner band line. Hence, the Doppler cores of these two fluid elements would not shield each other even if they contain substantial amounts of molecular hydrogen.

In other words, a typical fluid element of size $l_s$ in the molecular cloud can be strongly shielded by another fluid element only if it accidentally happens to fall at the same line-of-sight velocity. In turbulent gas with velocity dispersion $\Delta v$, the probability of another fluid element to be within the line width $b$ from any given fluid element is about $b/\Delta v$. Hence, a typical fluid element at depth $L$ inside a molecular cloud should be shielded by $dN_\H2/dv\sim N_\H2/\Delta v\sim \langle n_\H2\rangle_s L/\Delta v$, where $\langle n_\H2\rangle_s$ is the average molecular hydrogen density on a sonic scale and $\Delta v = b (L/l_s)^\gamma$ corresponds to the velocity width at depth $L$. For the Larson law with the slope $\gamma\approx 0.5$ that translates into a shielding column density which is the geometric mean of the total column density at depth $L$ and the column density on sonic scale, $N_\H2 \sim \langle n_\H2\rangle_s (l_sL)^{1/2}$, which is much less than the total column density through the cloud, since $l_s\ll L$.

However, there is a major flaw in this argument. Absorption lines are narrow only at low column densities, before the damping wings become important. At sufficiently large column densities, absorption lines in the Lyman and Werner bands become broad enough to begin to overlap \citep{bd77,ism:db96}, effectively rendering relative velocity shifts between different fluid elements unimportant. In other words, at sufficiently large column densities line radiative transfer in the Lyman and Werner bands behaves as continuum radiative transfer, and the effective length over which the column density is accumulated becomes the size of the whole cloud.

In Equation (\ref{eq:shfac}) the line overlap is described by the second term. To account for the supersonic turbulence inside the molecular cloud, Equation (\ref{eq:shfac}) can be modified by introducing two variables, $x_1$ and $x_2$, in place of $x$ as
\begin{equation}
S_{\H2}(N_\H2) = \frac{0.965}{(1+x_1/b_{5})^2} + \frac{0.035}{\sqrt{1+x_2}}\exp\left(-\frac{\sqrt{1+x_2}}{1180}\right),
\label{eq:lofac}
\end{equation}
where $x_{1} = (N_\H2 N_s)^{1/2}/5\times10^{14}\dim{cm}^{-2}$, $x_2 = N_\H2/5\times10^{14}\dim{cm}^{-2}$, $N_s = \langle n_\H2\rangle_s l_s$ is the column density on the sonic scale, and $N_\H2$ is the total $\H2$ column density, $N_\H2\approx \langle n_\H2\rangle_L L$. 
Below we will refer to this as the ``non-local'' line overlap treatment.

Line overlap becomes substantial at column densities $N_\H2\ga \mbox{a few}\times 5\times10^{14}\dim{cm}^{-2}\times 1180^2 \sim \mbox{a few}\times 10^{21}\dim{cm}^{-2}$, which is comparable to column densities at which shielding by dust is important in environments that have 10-20\% solar metallicity. Hence, in low metallicity environments the dust shielding becomes completely subdominant to the self-shielding of molecular hydrogen.

\section{Modeling Molecular Hydrogen on Galactic Scales}
\label{sec:gals}

Equation (\ref{eq:lofac}) applies to a given location in the molecular cloud. Modern cosmological or galactic scale simulations may not resolve molecular clouds at all or may resolve them only down to parsec scales. Hence, it is unlikely that Equation (\ref{eq:lofac}) can be used directly. Instead, we can imagine whole space being tessellated into regions (say, simulation cells - not necessarily all of the same size), some of which include pieces of molecular clouds. Each such region $j$ has a full distribution of column densities inside it, $\phi_j(N)$ (that samples both different locations inside the region and different directions at a given location).
 Hence, the average shielding factor is 
\[
  \left\langle S_\H2\right\rangle_j = \int S_\H2(N) \phi_j(N) dN,
\]
which, by the first mean value theorem for integration, can be represented as
\[
  \left\langle S_\H2\right\rangle_j = S_\H2(N_{\rm eff}) \int \phi_j(N) dN  = S_\H2(N_{\rm eff})
\]
(since $\phi_j$ is normalized to unity by definition). If the distribution $\phi_j$ was known, one could compute the effective column density $N_{\rm eff}$, but, at present, there are no models that attempt to determine $\phi_j$. Hence, we need to come up with an ansatz for $N_{\rm eff}$. 

Following \citet[][hereafter GK11]{ng:gk11} approach, we adopt a simple ``Sobolev-like" \emph{ansatz}
\[
  N_{\rm eff} \approx n_{\H2, j} L_{\rm Sob},
\]
where
\[
  L_{\rm Sob} = \frac{\rho}{2|\nabla\rho|}
\]
is calibrated from exact ray-tracing calculations (see GK11). With such an approximation the complete set of equations is obtained. 

In order to explore the effect of line overlap on galactic scales, we follow the methodology described in GK11.  We refer the reader to that paper for full details; here we only mention that we use the Adaptive Refinement Tree (ART) code \citep{misc:k99,misc:kkh02,sims:rzk08} to follow a region of the universe containing a couple dozen of galaxies of various masses up to $3\times10^{11}\Msun$ at $z\sim3$ with the mass resolution of $1.3\times10^6\Msun$ and peak spatial resolution of $260$ comoving pc ($65\dim{pc}$ in physical units at $z=3$). In order to explore the environmental dependence of the atomic-to-molecular transition, we run the simulations in the ``fixed ISM'' mode, in which we impose a 91.2--111~nm radiation field in and a fixed dust-to-gas ratio throughout the computational domain. 

We parametrize the 91.2--111~nm interstellar radiation field and the dust-to-gas ratio in our "fixed ISM" simulations in units of their values in the Milky Way so that $\U=1$ and $\D=1$ corresponds to the Milky Way ISM, $\U=10$ and $\D=0.5$ corresponds to the ISM conditions with 10 times higher radiation field and half the Milky Way dust-to-gas ratio, etc.

\begin{figure*}[t]
\includegraphics[width=0.5\hsize]{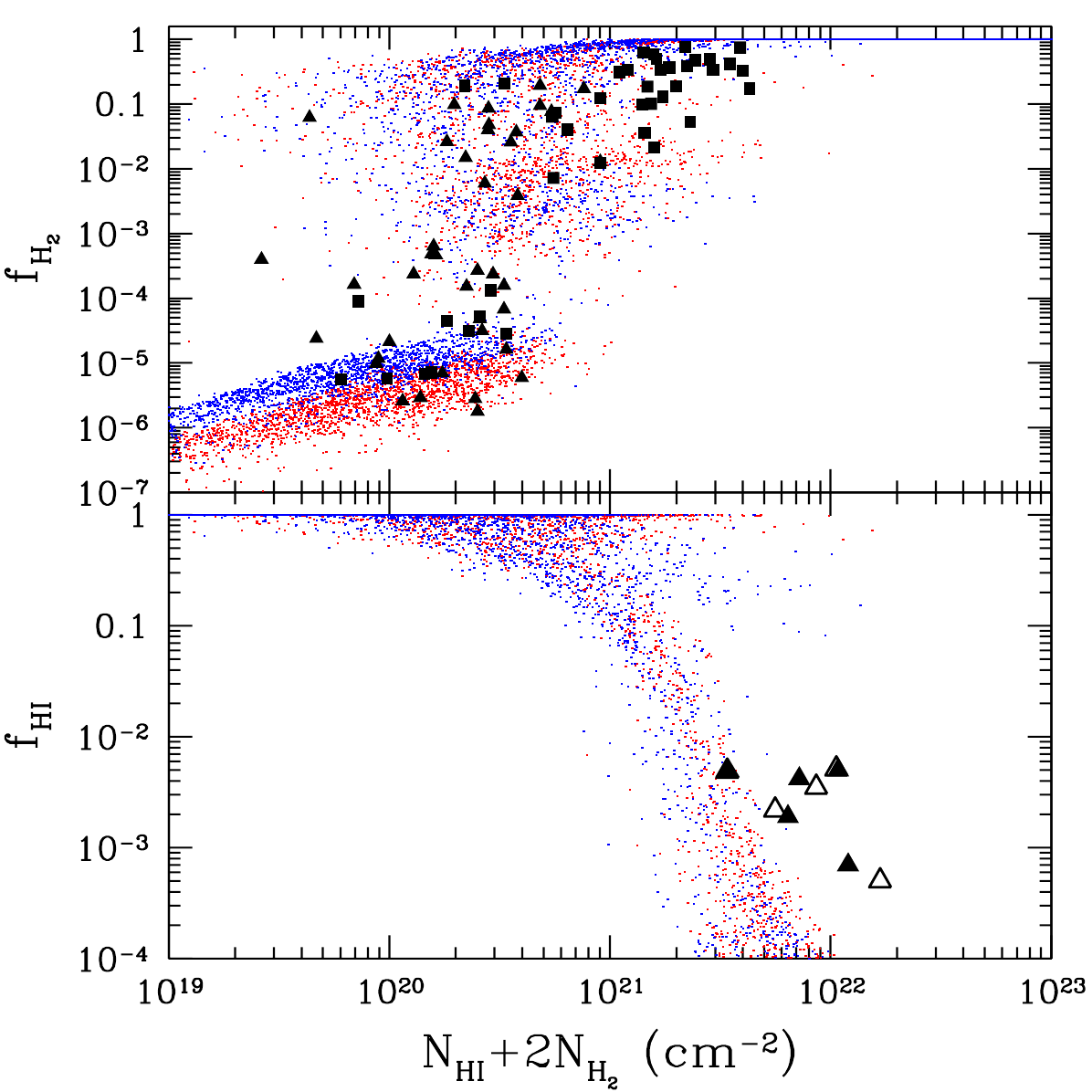}%
\includegraphics[width=0.5\hsize]{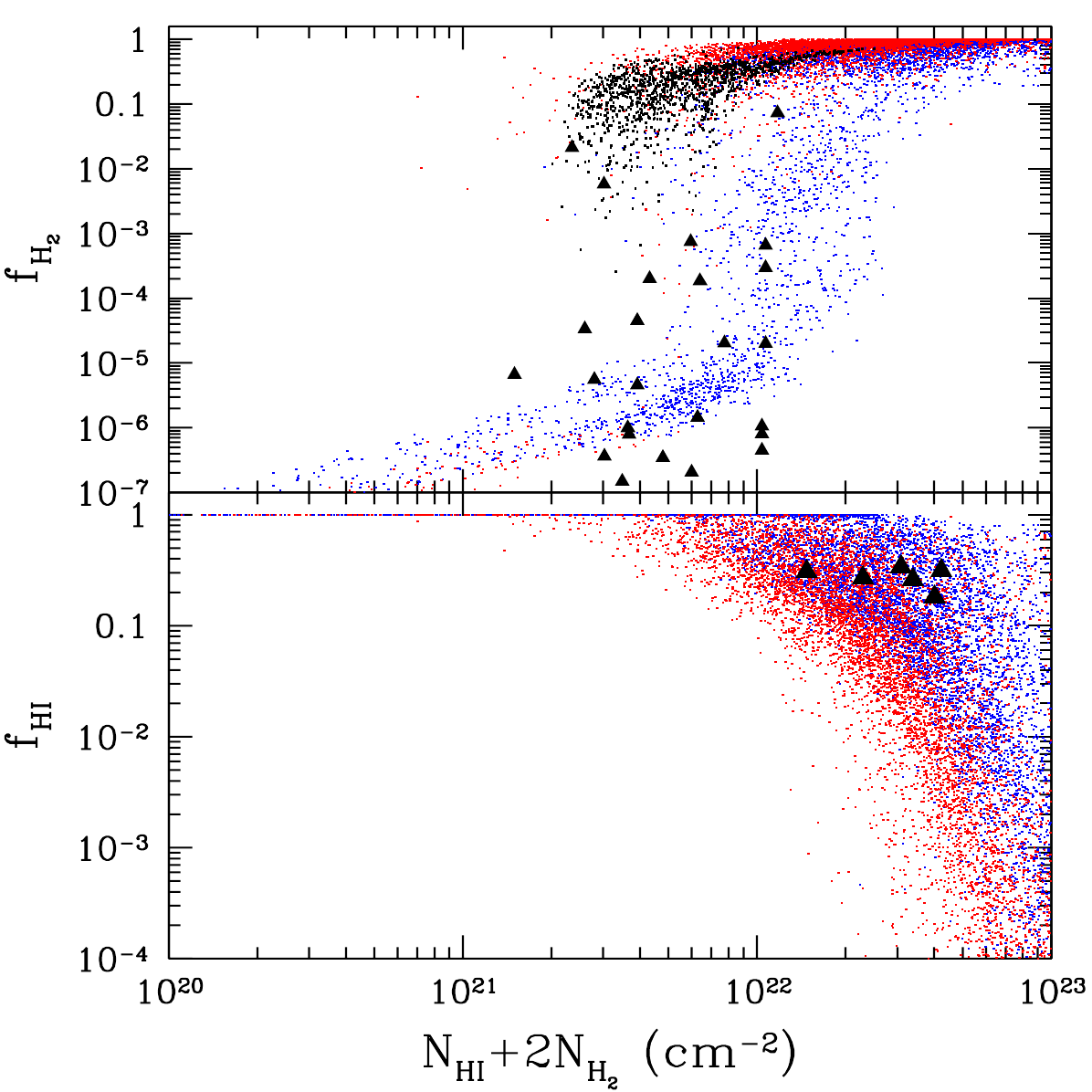}
\caption{Atomic (bottom) and molecular (top) gas fractions as functions of the total (neutral) hydrogen gas column density along individual lines of sight through the galactic disks. Colored points shows the simulation results, while black points are observations. The left panel shows the $(\D=1,\U=1)$ simulation case for $C_\rho=10$ (red) and $C_\rho=30$ (blue). Observational measurements of molecular fractions in the Milky Way galaxy from \citet{ism:gstd06} (filled triangles) and \citet{ism:wthk08} (filled squares) and atomic fractions measurements from \citet{ism:gl05}. The right panel shows $(\D=0.15,\U=10,C_\rho=10)$ simulation case as a possible model for the Small Magellanic Cloud. Filled triangles on the top panel show the measurements for SMC molecular fractions from \citet{ism:tsrb02}, open squares are the observational compilation of \citet{wxw12}, and black points are data from \citet{sfr:bljo11}. On the bottom panel the measurements are from \citet{ism:lbsm07}. Red points show our complete model, while blue points are for the simulation in which line overlap is ignored (i.e.\ using the power-law approximation of Eq.\ \ref{eq:nolo} for the self-shielding factor). As one can see, the SMC data are in better agreement with the model that includes line overlap.\label{fig:calib}}
\end{figure*}

The complete description of the numerical model for formation of molecular hydrogen is presented in the Appendix of GK11. The only changes to that model that we use in this paper are (i) cooling and heating of the gas are modeled with more physically realistic cooling and heating functions of \citet{ng:gh12}, and, most importantly, (ii) the self-shielding factor for molecular hydrogen $S_\H2$ is determined from Equation (\ref{eq:lofac}) above rather than from Equation (A11) of GK11, which, by using a power-law approximation to the self-shielding factor,
\begin{equation}
  S_\H2 \approx \left\{
  \begin{array}{ll}
    1, & \mbox{for } N_\H2 < 10^{14}\dim{cm}^{-2}, \\
    \left(N_\H2/10^{14}\dim{cm}^{-2}\right)^{-3/4}, & \mbox{for }
    N_\H2 > 10^{14}\dim{cm}^{-2}, 
  \end{array}
  \right.
  \label{eq:nolo}
\end{equation}
does not explicitly account for line overlap (the exponential factor in Eqs.\ \ref{eq:shfac} and \ref{eq:lofac}).

The $\H2$ formation model dependents on two parameters - the clumping factor $C_\rho \equiv \langle n_\Ht^2\rangle_{\rm cell}/\langle n_\Ht\rangle_{\rm cell}^2$ of gas inside a simulation cell (which is not resolved, and, hence, needs to be parameterized) and the sonic length $l_s$. The good news is that line overlap effectively eliminates any dependence on $l_s$, so, in practice, just one parameter, the clumping factor, matters.

\section{Calibration of the $\H2$ Formation Model}

While one can come up with reasonable estimates for the model parameters, the ultimate parameter choice is dictated by comparison with observations, given that our model is essentially an empirical one. For calibrating the model we use two types of observational data: measurements of gas fractions along lines of sight to individual stars for atomic \citep{ism:gl05} and molecular gas in the Milky Way and Magellanic Clouds \citep{ism:tsrb02,ism:gstd06,ism:wthk08,sfr:bljo11} and the measurements of atomic and molecular gas surface densities in nearby spirals from \citet{misc:wb02}.

\begin{figure}[b]
\includegraphics[width=1\hsize]{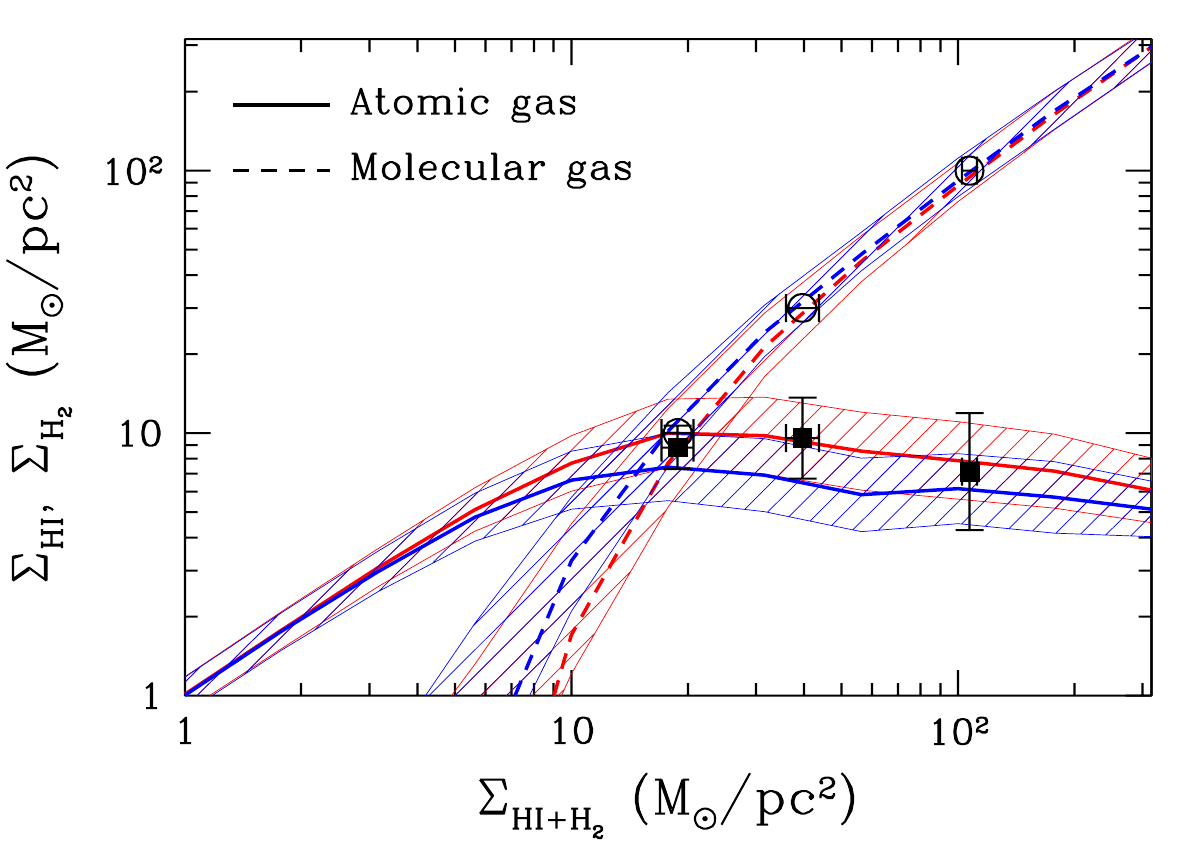}%
\caption{Average atomic and molecular gas surface densities as functions of the total (neutral) hydrogen gas surface density averaged over $500\dim{pc}$ scale for the $(\D=1,\U=1)$ simulation case for $C_\rho=10$ (red) and $C_\rho=30$ (blue, lines/bands for mean/rms). Filled squares and open circles with error bars mark the observed average atomic and molecular hydrogen surface densities at $\Smol=10$, $30$, and $100\Msun\dim{pc}^{-2}$ from \citet{misc:wb02}. The error-bars on the observational points show the dispersion around the average rather than the error of the mean.\label{fig:wong}}
\end{figure}

The best-fit $\H2$ formation model is shown in Figures \ref{fig:calib} and \ref{fig:wong} for two values of the clumping factor $C_\rho=10$ and $30$. Both values provide acceptable fits to the data; the $C_\rho=10$ case fits the \citet{misc:wb02} measurement perfectly, while the $C_\rho=30$ is slightly better for the translucent clouds. The model is insensitive to the value of the sonic length $l_s$ as long as $l_s<10\dim{pc}$.

One can reasonably wonder whether the calibration data we use actually test the effect of line overlap. The Milky Way data and \citet{misc:wb02} galaxies have too high metallicities for the line overlap to be important. Hence, only the SMC data have any discriminating potential for models with different treatments of line overlap. To verify that, we show in the right panel of Figure \ref{fig:calib} also an SMC-like model that neglects line overlap (i.e. uses Eq.\ \ref{eq:shfac} instead of Eq.\ \ref{eq:lofac} for modeling $\H2$ self-shielding). 

For our best-fit parameters, the model without line overlap is clearly disfavored by the SMC data. One can re-calibrate the model and make the no-line-overlap model fit the SMC data better, but such a model would then fail the Milky Way calibration. Hence, the combination of Milky Way and the SMC constraints appears to favor the model with non-local line overlap treatment, thus justifying our approach. 

One can notice, though, that GK11 did manage to reach a reasonable match to both the Milky Way and the SMC calibration data without line overlap. That apparent discrepancy is due to the fact that GK11 had no access to \citet{sfr:bljo11} data (small black points on the left panel of Fig.\ \ref{fig:calib}), but only to the older data from \citet{ism:tsrb02}. The latter data set is indeed in an acceptable agreement with the blue points from the left panel of Fig.\ \ref{fig:calib}. Hence, the measurements of \citet{sfr:bljo11} are the only data set that prefers the model with non-local line overlap over that without it.

\section{Results}
\label{sec:res}

\begin{figure}[t]
\includegraphics[width=1\hsize]{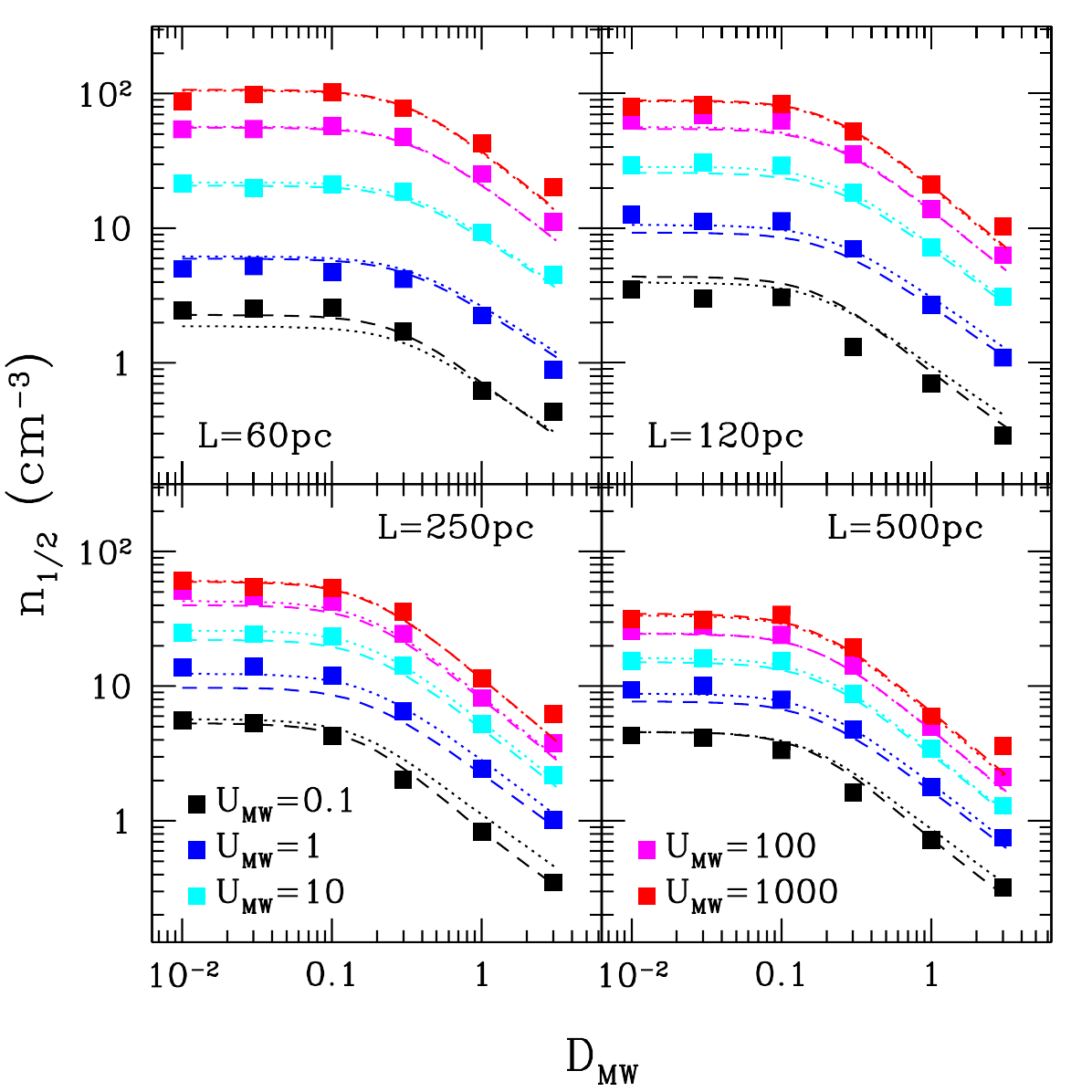}
\caption{Average total hydrogen number density of atomic-to-molecular
  gas transition (defined as $f_\H2=1/2$) as a function of the
  dust-to-gas ratio $\D$ and the interstellar UV radiation field $\U$
  for all our simulations at four values of the averaging spatial
  scale. This plot can be compared with Fig.\ 2 of GK11. Solid lines
  show fitting formula of Equation (\ref{eq:n12}), while dotted lines on the first panel are GK11 fits without line overlap (using the power-law approximation of Eq.\ \ref{eq:nolo} for the self-shielding factor).\label{fig:nthfit}}
\end{figure}

With the complete model in hand, we repeat the procedure of GK11:
sampling the interstellar UV radiation field strength and dust-to-gas
ratios over a grid of values to investigate the dependence of the
atomic-to-molecular hydrogen transition on these two physical
parameters.

Figure \ref{fig:nthfit} shows the first main result of this paper - the dependence of the characteristic density of the atomic-to-molecular hydrogen transition $n_{1/2}$ (defined as the density at which the mean $\H2$ fraction reaches 50\%) on the interstellar UV radiation field and the dust-to-gas ratio. It can be compared directly to Fig.\ 2 of GK11.

The primary feature of Fig.\ \ref{fig:nthfit} is the saturation of $n_{1/2}$ as a function of the dust-to-gas ratio for $\D\la0.1$ - this is a direct consequence of line overlap, which dominates over dust shielding for low dust abundances. At extremely low dust abundances ($\D\la10^{-3}$) the time-scale for molecular hydrogen formation (which is proportional to the dust abundance) becomes long compared to other galactic time-scales (rotation period, accretion time-scale, etc) and the situation becomes more complex \citep{ism:k12}. In addition, the time-scale for dust formation also becomes long \citep{ism:d09,ism:i13}, and in such a dynamical environment there is no alternative to full time-dependent non-equilibrium treatment.

At higher dust-to-gas ratios considered here the atomic-to-molecular transition is well-behaved, and a useful approximation can be developed as an alternative to the detailed calculation of molecular chemistry.

\begin{figure}[t]
\includegraphics[width=1\hsize]{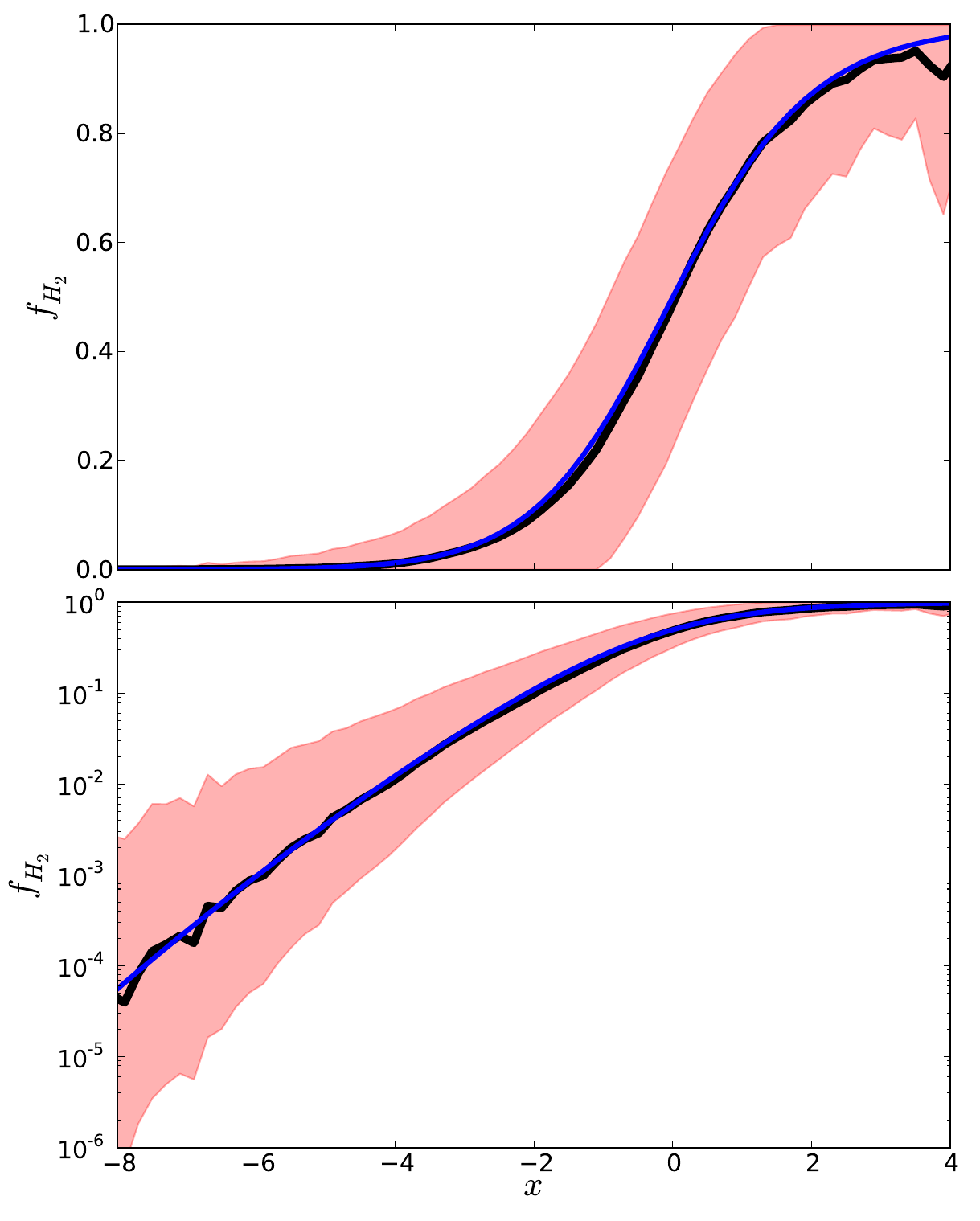}
\caption{Molecular hydrogen fraction as a function of variable $x$ (Equation \ref{eq:xdef}). The black line shows the approximation average $f_\H2$ for each value of $x$. The pink band shows the full rms scatter
of $f_\H2$ as a function of $x$. The blue line tracks the approximation of Equation (\ref{eq:fh2}).}
\label{fig:nfit}
\end{figure}

Following the approach of GK11, we can fit $n_{1/2}$ as a function of $\U$, $\D$. In addition, we also consider the effect of averaging our numerical results over a larger spatial scale $L$ - that would be useful if these fits are to be used in lower resolution simulations. 

It is important to note, though, that the spatial averaging of a high resolution simulation is \emph{different} from using the full chemical model in a lower resolution simulation. Namely, as GK11 noted, the full chemical model is reasonably resolution-independent, and can be used in simulations with spatial resolutions better than about $130\dim{pc}$, while coarser resolved simulations should not use the full chemical model at all, and that statement is not affected by our account of the line overlap. If, however, a simulation is designed to use our fitting formulae, then the simulation resolution needs to be explicitly accounted for, since spatial averaging is explicitly dependent on the averaging scale. Consider, for example, a simulation with spatial resolution of, say, $260\dim{pc}$. One can imagine that each $(260\dim{pc})^3$ cell in such a simulation contains $4^3$ of $65\dim{pc}$ cells we use here. Hence, in order to provide atomic and molecular abundances in each cell of the low resolution simulation, we should sum up atomic and molecular abundances over $4^3$ of our high resolution cells. Since, in the low resolution simulation, the physical quantities are only defined in $(260\dim{pc})^3$ cells, we would like to express the total atomic and molecular abundances in our $4^3$ cells as functions of atomic and molecular abundances in the low resolution simulation; such a parameterization will not, of course, be exact, and will explicitly depend on the averaging scale.

For our highest resolution of $65\dim{pc}$ we use the simulated values directly. For larger spatial scales $L$
we average simulation cells in groups of 8, progressively sampling the scales
$L=130\dim{pc}$, $260\dim{pc}$, etc. The following fitting formulae describe our simulated results reasonably well (we maintain the functional dependence used by GK11 as it is justified on physical grounds):
\begin{equation}
  n_{1/2} = n_*\frac{\Lambda}{g},
  \label{eq:n12}
\end{equation}
where 
\begin{eqnarray}
  \Lambda & = & \mbox{ln}\left(1+\left(0.05/g+\U\right)^{2/3}g^{1/3}/U_*\right), \nonumber\\
  g & = & \left(\D^2+D_*^2\right)^{1/2}, \nonumber
\end{eqnarray}
and
\begin{eqnarray}
  n_* & \equiv & 14\dim{cm}^{-3}\frac{D_*^{1/2}}{S},\nonumber\\
  U_* & \equiv & 9\frac{D_*}{S},\nonumber\\
  D_* & = & 0.17\frac{2+S^5}{1+S^5}, \nonumber
\end{eqnarray}
and $S \equiv L/100\dim{pc}$.

With the known value of $n_{1/2}$, the full simulations results for the average molecular fraction as a function of total hydrogen density $n_\Ht$ can be fitted with the following simple expression,
\begin{equation}
  \langle f_\H2\rangle(n_\Ht) = \frac{1}{1+\exp\left(-x(1-0.02x+0.001x^2)\right)},
  \label{eq:fh2}
\end{equation}
where
\begin{equation}
  x = w \, \mbox{ln}\left(\frac{n_\Ht}{n_{1/2}}\right).
  \label{eq:xdef}
\end{equation}
and
\[
  w = 0.8 + \frac{\Lambda^{1/2}}{S^{1/3}}.
\]
The accuracy of this approximation is demonstrated in Figure \ref{fig:nfit}, together with the rms scatter $f_\H2$ as a function of $x$. Approximation (\ref{eq:fh2}-\ref{eq:xdef}) replaces equations (6-9) of GK11.

In numerical simulations parametrizing the molecular hydrogen abundance in terms of the gas density is most convenient, as the density is always directly followed in a simulation. For semi-analytical models a more appropriate quantity is the gas surface density. Often it is more convenient to use a ratio of molecular-to-atomic surface densities $R\equiv \Smol/\Shi$ instead of individual surface densities, both because the ratio is know to scale simply with galactic properties \citep{ism:br06} and because the ratio is easier to fit.

All of our simulations can be fitted on sufficiently large scales ($L\gtrsim500\dim{pc}$) with a simple power-law dependence of $R$ on the surface density of neutral gas,
\begin{equation}
  R \approx q\frac{1+\eta q}{1+\eta},~~q\equiv \left(\frac{\Sgas}{\Sigma_{R=1}}\right)^\alpha,
  \label{eq:sfit}
\end{equation}
where $\eta\approx0$ on kpc scale and $\eta=0.25$ on 500-pc scale,
\begin{equation}\label{eq:alpha}
  \alpha = 1 + 0.7\frac{s^{1/2}}{1+s},
\end{equation}
and the surface density at which the molecular and atomic fractions become equal
\begin{equation}\label{eq:R=1}
  \Sigma_{R=1} = \frac{40\Msun/\dim{pc}^2}{g}\frac{s}{1+s}
\end{equation}
with $s \equiv \left(0.001+0.1\U\right)^{0.7}$.

\begin{figure}[t]
\includegraphics[width=1\hsize]{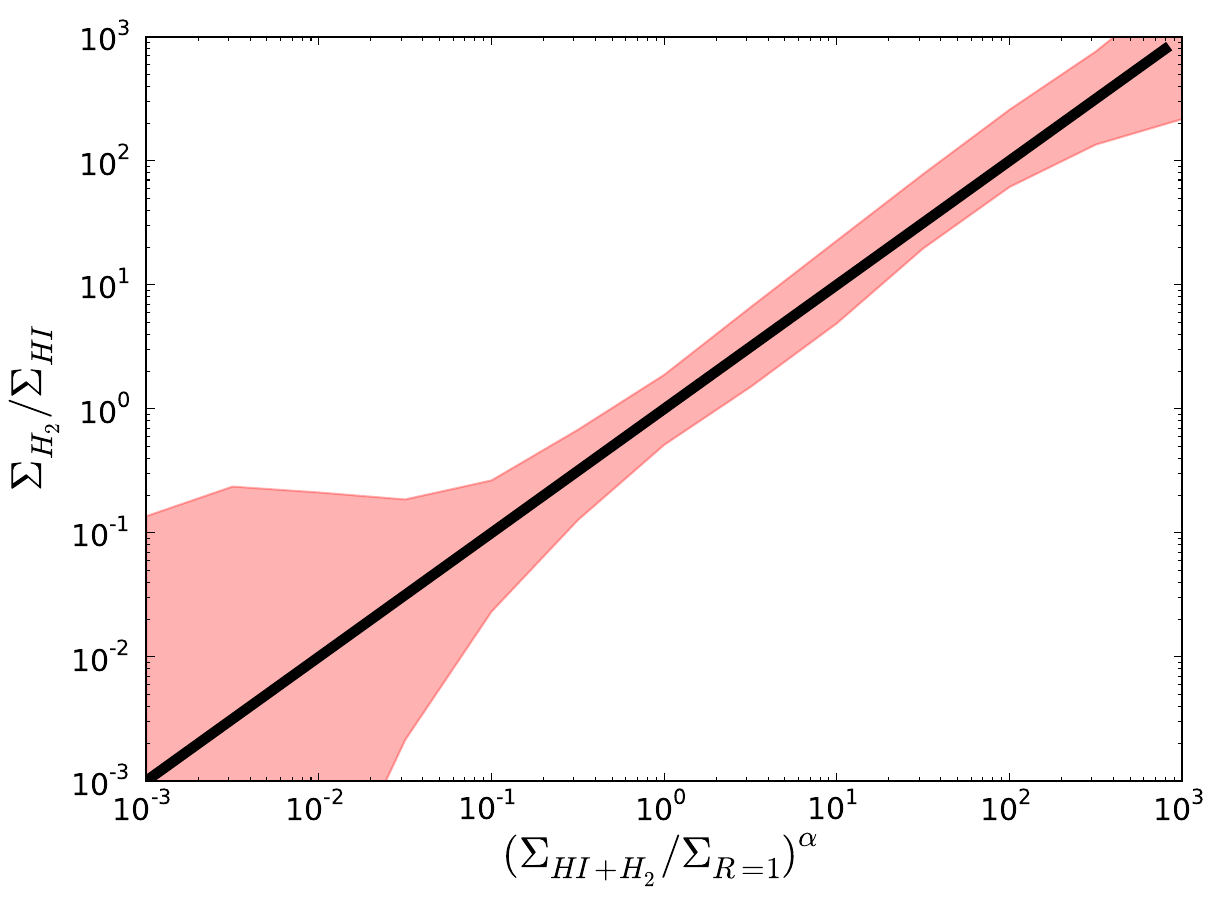}
\caption{Ratio of molecular to atomic hydrogen surface densities as a function of surface density of neutral hydrogen (Eq. \ref{eq:sfit}, black line), where $\alpha$ and $\Sigma_{R=1}$ are given by Eqs.\ (\ref{eq:alpha}) and (\ref{eq:R=1}). The pink shaded area has the same meaning as in Fig.\ \ref{fig:nfit}: the rms deviation around the mean relation at all spatial locations in all of our simulations. This plot can be compared with Fig.\ 7 of GK11.\label{fig:sthfit}, although the two figures are not completely analogous, since here we show the ratio of $\Smol/\Shi$ instead of just $\Smol$ in GK11.}
\end{figure}

Note, that in the limit of high column density the functional form for the atomic hydrogen surface density,
\[
  \Shi = \frac{\Sntr}{1+R},
\]
is not constant if $\alpha$ is not equal to 1. Hence, contrary to the assumption of GK11, the atomic surface density does not saturate at high densities, but may continue to decrease gradually. This behavior is, indeed, in better agreement with observational data of \citet{misc:wb02}, which show a variation in asymptotic behavior of $\HI$ surface density. 

Figure \ref{fig:sthfit} shows the accuracy of this approximation. The rms scatter around the mean relation is rather small at around $\Sigma_{R=1}$ but increases towards high and, even more significantly, toward low surface densities. For $R\la 0.1$ the approximation of Equation \ref{eq:sfit} effectively breaks down, as the scatter becomes too large.

\section{Conclusions}
\label{sec:cons}

We discuss the effect of line overlap in the Lyman and Werner bands on the atomic-to-molecular hydrogen transition on galactic scales. While in Milky Way like environments dust shielding is important and can not be neglected, line overlap significantly enhances self-shielding and makes it dominant over dust shielding in the environments with dust-to-gas ratios below about 10\% of the Milky Way value (i.e.\ in galaxies like SMC and smaller).

\begin{figure}[t]
\includegraphics[width=1\hsize]{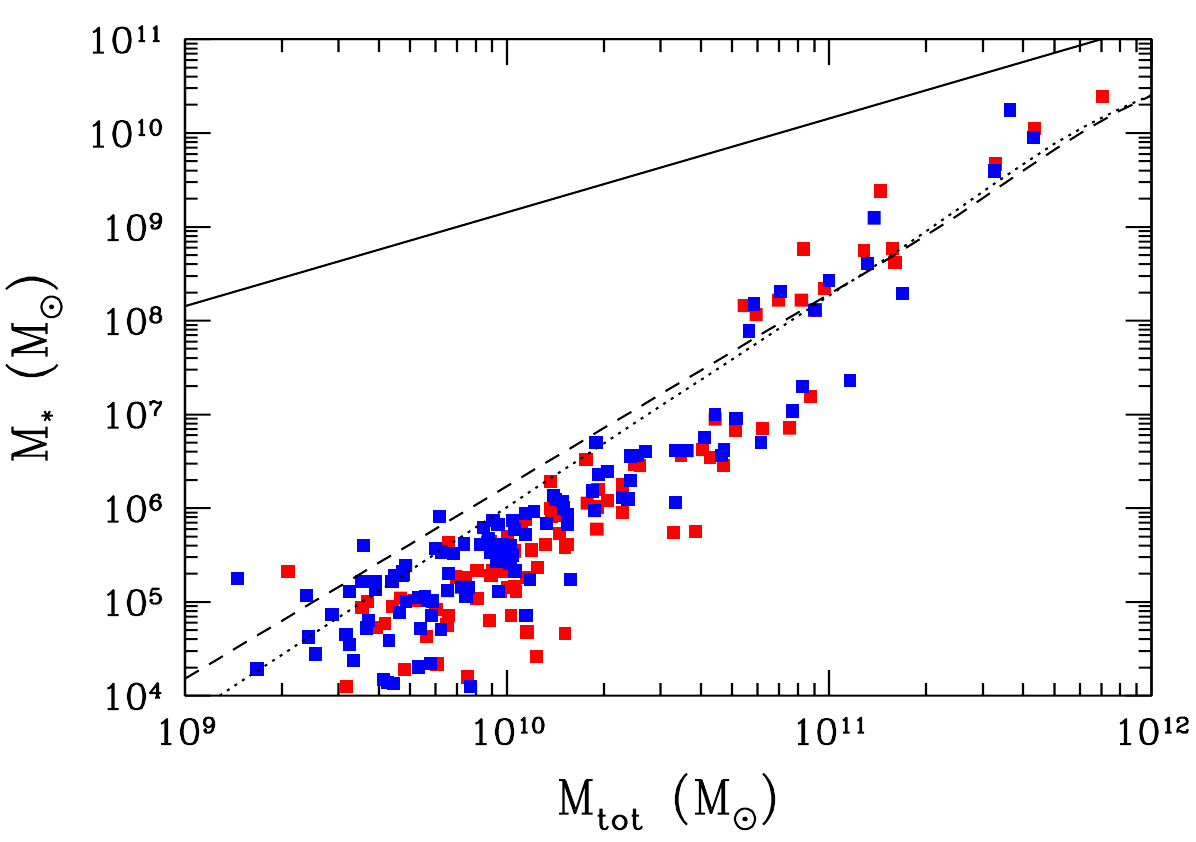}
\caption{Stellar vs total mass for galaxies simulated in illustrative simulations, one including non-local line overlap (red symbols) and another one with line overlap neglected (using the power-law approximation of Eq.\ \ref{eq:nolo} for the self-shielding factor - blue symbols), at $z=3$ and $z=2$ (both redshifts combined on this plot). Dotted and dashed lines show the $z=3$ and $z=2$ predictions for this relation from \protect\citet{gals:bwc13}, while the solid line traces the universal baryon fraction. While line overlap is important for determining the atomic-to-molecular gas transition, it appears to have a negligible effect on the overall properties of simulated galaxies.\label{fig:msmt}}
\end{figure}

So far all cosmological simulations that modeled formation of molecular hydrogen ignored line overlap. In particular, results of GK11 were found \citep{ng:kg11} to be fully consistent with the fitting formulae of \citet{sfr:kmt09a}. Since we find that line overlap makes a large effect on the atomic-to-molecular transition at dust abundances below about 10\% of the Milky Way value, models of this transition similar to GK11 and \citet{sfr:kmt09a} underestimate molecular abundance in extremely dust poor galaxies.

How important this effect is to the detailed physics of galaxy formation is currently unclear. For example, in an early galaxy with fully atomic ISM the formation of the very first molecular gas should still be controlled by dust shielding; however, having formed, line overlap makes molecular gas more resistant to photo-desctruction. Full exploration of the role of line overlap in galaxy formation is well beyond the scope of a single paper. Here we only present a simple illustrative example of the ``cosmo I'' simulation from GK11, with a single modification to the stellar feedback model. In the original GK11 simulation only the thermal energy feedback was implemented, and such implementation is known to be inefficient. In the simulations used here we adopt a currently widely used ``blastwave'' or ``delayed cooling'' feedback model \citep{sims:sdqg09,sims:gbmb10,sims:atm11,sims:bsgk12,sims:aklg13,sims:sbmw13}, with the delay time parameter set to $30\dim{Myr}$. Such a model for feedback is known to produce galaxies with overall properties (stellar masses, rotation curves, etc) in reasonable agreement with observations \citep{gals:mgb13}.

A stellar mass - total mass relation for two illustrative simulations - one with non-local line overlap included, and another one with line overlap neglected (using the power-law approximation of Eq.\ \ref{eq:nolo} for the self-shielding factor) - is presented in Figure \ref{fig:msmt}. Line overlap has only minor effect on stellar masses of simulated galaxies, and that result is not too surprising. After all, it is well established that stellar masses and star formation rates of galaxies are controlled by the feedback. Any increase in star formation due to line overlap is going to be offset by stronger feedback on the time-scale of several tens of Myr, and the average, long-term efficiency of star formation will remain at the ``self-regulated'' value \citep[c.f.][]{sims:hqm11,sims:aklg13,sims:hko13}.

\acknowledgements

We are grateful to Andrey Kravtsov, Dan Welty, Amiel Sternberg, and an anonymous referee for valuable comments and suggestions that significantly improved the original manuscript.

Fermilab is operated by the Fermi Research Alliance, LLC, under
contract No. DE-AC02-07CH11359 with the United States Department of
Energy. NYG work was also supported in part by the NSF grant
AST-1211190. BTD was supported in part by NSF grant AST-1008570.

\bibliographystyle{apj}
\bibliography{ms,ng-bibs/self,ng-bibs/ism,ng-bibs/misc,ng-bibs/sims,ng-bibs/sfr,ng-bibs/gals}

\end{document}